\documentclass[prl,twocolumn,twoside,superscriptaddress,showpacs,floatfix]{revtex4}
\usepackage{amsmath,graphicx,multirow,natbib,hyperref}
\hypersetup{colorlinks,citecolor=black,filecolor=black,linkcolor=black,urlcolor=black}

\begin{document}

\title{Correlated disorder in the Kuramoto model: \\ 
Effects on phase coherence, finite-size scaling, and dynamic fluctuations}

\author{Hyunsuk Hong}
\email{hhong@jbnu.ac.kr}
\affiliation{Department of Physics and Research Institute of Physics and Chemistry, Chonbuk National University, Jeonju 561-756, Korea}

\author{Kevin P. O'Keeffe}
\email{kpo24@cornell.edu}
\affiliation{Center for Applied Mathematics, Cornell University, New York 14853, USA} 

\author{Steven H. Strogatz}
\email{shs7@cornell.edu}
\affiliation{Department of Mathematics, Cornell University, New York 14853, USA} 

\date{\today}

\begin{abstract}
We consider a mean-field model of coupled phase oscillators with quenched disorder in the natural frequencies and coupling strengths. A fraction $p$ of oscillators are positively coupled, attracting all others, while the remaining fraction $1-p$ are negatively coupled, repelling all others. The frequencies and couplings are deterministically chosen in a manner which correlates them, thereby correlating the two types of disorder  in the model. We first explore the effect of this correlation on the system's phase coherence. We find that there is a a critical width $\gamma_c$ in the frequency distribution below which the system spontaneously synchronizes. Moreover, this $\gamma_c$ is independent of $p$. Hence, our model and the traditional Kuramoto model (recovered when $p=1$) 
have the same critical width $\gamma_c$. We next explore the critical behavior of the system by examining the finite-size scaling and the dynamic fluctuation of the traditional order parameter. We find that the model belongs to the same universality class as the Kuramoto model with 
deterministically (not randomly) chosen natural frequencies for the case of $p<1$.
\end{abstract}

\maketitle


\section{I. Introduction}

Since its inception in 1975, the Kuramoto model \cite{kuramoto84} has been applied to a wide variety of synchronization phenomena \cite{winfree, strogatz00, pikovsky03, acebron05, rodrigues15}, including  arrays of Josephson junctions \cite{josephson, josephson1}, electrochemical oscillations \cite{chemical_osc, chemical_osc1}, the dynamics of 
power grids~\cite{dorfler14,rodrigues15}, and even rhythmic applause~\cite{clapping}.

In the original model, Kuramoto considered oscillators with distributed natural frequencies, coupled all-to-all with constant strength. But in many biological systems this coupling scheme is unrealistic. For example, in neuronal networks and in coupled $\alpha$, $\beta$, and $\delta$-cells in the 
pancreatic islets~\cite{Borgers03,Menge11,Hellman09}, the oscillators are 
coupled to each other both positively and negatively. Many researchers have modified the Kuramoto model to include such interactions of mixed sign, leading to new 
macroscopic phenomena~\cite{Zanette05, HS11, van_hemmen, iatsenko13,iatsenko14} such as the traveling wave state, the $\pi$ state, and the mixed state. 
There has also been some evidence of glassy dynamics \cite{glass1,glass2_glass3}.

These rich phenomena arise from the competition between the positive coupling, the negative coupling, and the distributed frequencies. In this paper, we include another element in the competition: correlations. We consider a simple mean-field model in which mixed couplings exist and are correlated with the natural frequencies. We recently studied such a system in Ref.~\cite{HKS16}. There, we considered \emph{equal} numbers of positively and negatively coupled oscillators with frequencies distributed according to a Lorentzian with zero mean and variable width. We showed that even though the mean frequency and mean coupling were both zero, a transition from incoherence to partial synchrony was possible. This was somewhat of a surprise: one might expect that since the average coupling is zero, the system is effectively uncoupled, and therefore only incoherence should be possible. 

We were curious to see if correlated disorder would give other surprises when the average coupling was non-zero. To this end, we here extend the analysis in Ref.~\cite{HKS16} to incorporate \emph{variable} numbers of positively and negatively coupled oscillators. We also study the critical behavior near the transition from incoherence to partial synchrony, by investigating the finite-size scaling and the dynamic fluctuation of the order parameter. 

The paper is organized as follows. In Section II, we define our model and its order parameter, and  specify the correlations. We present our theoretical analysis in Section III, and compare these predictions with numerical results in Section IV.  In Section V, we study the finite-size scaling behavior of the order parameter, and then explore its dynamic fluctuation in Section VI. Lastly, we provide a brief summary in Section VII.


\section{II. The  Model}
The governing equation for our model is
\begin{equation}
\frac{d\phi_i}{dt} = \omega_i + \frac{1}{N} \sum_{j=1}^{N} \xi_j \sin(\phi_j-\phi_i),
\label{eq:model}
\end{equation}
\noindent
for $i = 1, \dots, N$. Here $\phi_i$, $\omega_i$ and $\xi_i$ are the phase, natural frequency, and coupling of oscillator $i$. The $\omega_i$ are drawn from a Lorentzian distribution with center frequency $\langle \omega \rangle$ and width $\gamma$. By going to a suitable rotating frame, we can set $\langle \omega \rangle = 0 $ without loss of generality, giving
\begin{equation}
g(\omega)= \frac{\gamma}{\pi} \frac{1}{\omega^2 + \gamma^2}.
\label{eq:g_omega}
\end{equation}
\noindent
We interpret the distributed frequencies as a kind of \textit{disorder}. This is because, as we will show, a spread in frequencies inhibits synchrony: the wider the distribution, the less synchronous the population. We also make the couplings $\xi$ disordered. For simplicity, we draw them from a double-$\delta$ distribution function,

\begin{equation}
\Gamma(\xi) = p\delta(\xi-1)+(1-p)\delta(\xi+1),
\label{eq:Gammaxi_delta}
\end{equation}
\noindent
so that a fraction $p$ of oscillators have positive coupling. Positively coupled oscillators are ``social''; they attract other oscillators, which promotes synchrony. The remaining fraction $1-p$ of oscillators have negative coupling. They are ``antisocial,'' tending to repel others, which inhibits synchrony.

We have previously studied oscillators with $\omega_i$ and $\xi_i$ 
distributed according to \eqref{eq:g_omega} and \eqref{eq:Gammaxi_delta} 
in Ref.~\cite{HS12}.  
In that work, the $\omega_i$ and $\xi_i$ were independent; the two types of disorder were uncorrelated. Surprisingly, the system behaved in the same way as the 
traditional Kuramoto model: the oscillators switched from incoherence to partial synchrony through a second-order phase transition. 

But what if the disorder is correlated? Are there new phenomena? To explore these questions, we correlate  $\{ \omega_i \}$ and $\{ \xi_i \}$ as follows. First, we choose the ${\omega_i}$ {\it{deterministically}} such that their cumulative distribution function matches that  implied by $g(\omega)$. This condition yields the deterministic frequencies ${\omega_i}$ as the solutions of 
\begin{equation}
\frac{i-0.5}{N}=\int_{-\infty}^{\omega_i} g(\omega) d\omega,
\label{eq:reg_process_gomega}
\end{equation}
\noindent
for $i = 1, \ldots, N$.  For the particular case of the Lorentzian distribution assumed here,  this procedure yields
\begin{equation}
\omega_i = \gamma \tan\Bigg[\frac{i \pi}{N}-\frac{\pi (N+1)}{2N}\Bigg],~~~i=1,\ldots,N. 
\label{eq:regLorw}
\end{equation}
\noindent

\begin{figure}[!htpb]
        \includegraphics[width=0.9\linewidth]{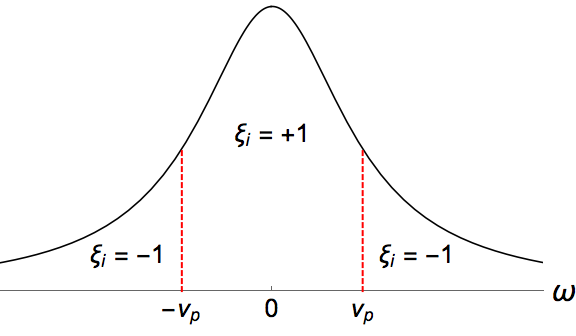}
        \caption{\label{fig:symmetric_disorder} (Color Online) Symmetrically correlated disorder.  Oscillators with $\omega_i < -\nu_p$ and $\omega_i > \nu_p$ have $\xi_i =-1$, and those with $-\nu_p < \omega_i < \nu_p$ have $\xi_i = +1$. The frequency $\nu_p$ is given by Eq.~\eqref{nu_p}, which we derive in the Section III, subsection ``Branch 1''. 
}
\end{figure}

Next, we deterministically choose the couplings according to
\begin{equation}
\xi_i  =
\left\{
\begin{array}{ll}
-1, & i=1, \ldots, \frac{(1-p)N}{2},\\
+1, & i=\frac{(1-p)N}{2}+1, \ldots, \frac{(1+p)N}{2}, \\
-1, & i=\frac{(1+p)N}{2}+1, \ldots, N,
\end{array}
\right.
\label{eq:xi_sym} 
\end{equation}
\noindent
which is shown in Fig.~\ref{fig:symmetric_disorder}.  In this way, $\omega_i$ 
is ``symmetrically'' correlated with $\xi$, in the sense that there are equal numbers of positively and negatively coupled oscillators distributed about $\omega = 0$.

\section{III. Analysis}

Phase coherence is conveniently measured by the complex 
order parameter $Z$, defined as~\cite{kuramoto84} 
\begin{equation}
Z\equiv R e^{i\Theta} = \frac{1}{N}\sum_{j=1}^N e^{i\phi_j}, 
\label{eq:Z}
\end{equation}
where $R$ and $\Theta$ measure the phase coherence $(0 \leq R \leq 1)$ and the average phase, respectively. 
We also consider another order parameter defined by
\begin{equation}
W\equiv S e^{i\Phi} = \frac{1}{N}\sum_{j=1}^N \xi_j e^{i\phi_j}, 
\label{eq:W}
\end{equation}
which is a sort of {\it{weighted}} mean field~\cite{HS12,HKS16}.  This order parameter 
lets us rewrite Eq.~\eqref{eq:model} as 
\begin{equation}
\dot\phi_i = \omega_i-S\sin(\phi_i-\Phi).
\label{eq:dpdtwithW}
\end{equation}

We now analyze the stationary states of our system using Kuramoto's classic self-consistency analysis \cite{kuramoto84}. In these states, the macroscopic variables $R, \Theta, S$ and $\Phi$ are all constant in time. Consequently, the oscillators governed by Eq.~(\ref{eq:dpdtwithW}) are divided into two types. The oscillators with  $|\omega_i| \le S$ become {\it{locked}}, having  stable fixed points  given by $\phi_i^*=\Phi+\sin^{-1}(\omega_i/S)$.  The other oscillators with $|\omega_i| > S$, on the other hand, are {\it{drifting}} with nonzero phase 
velocity ($\dot\phi_i\neq 0$). They rotate nonuniformly, having the stationary density 
\begin{equation}
\rho(\phi,\omega)=\frac{\sqrt{\omega^2-S^2}}{2\pi|\omega-S\sin(\phi-\Phi)|},
\end{equation}
found from requiring $\rho \propto 1/\dot{\phi}$ and imposing the normalization condition $\int \rho \ d \phi = 1$ for all $\omega$.  

This splitting of the population into locked and drifting subpopulations lets us find the stationary values of $W$ via the self-consistency equation
\begin{equation}
W=\langle \xi e^{i \phi} \rangle = \langle \xi e^{i \phi} \rangle_{\rm{lock}} + \langle \xi e^{i \phi} \rangle_{\rm{drift}} .
\label{eq:self_consistency_eqn}
\end{equation}
We can set $\Phi = 0$ without loss of generality, by the rotational symmetry of the system, so that $W = S e^{0} = S$. This leads to
\begin{equation}
S=\langle \xi e^{i \phi} \rangle = \langle \xi e^{i \phi} \rangle_{\rm{lock}} + \langle \xi e^{i \phi} \rangle_{\rm{drift}} .
\label{eq:self_consistency_eqn1}
\end{equation}
As we vary the fraction of positively coupled oscillators $p$ and the width of Lorentzian distribution $\gamma$, the solution set for $S$ will have two branches. The first branch occurs when only the oscillators with $\xi_i > 0$ are locked. The second branch occurs when both $\xi_i >0$ and $\xi_i <0$ are locked.

\subsection{Branch 1}

To solve for the first branch, we start by determining the maximum frequency of the locked oscillators with $\xi_i > 0$. Let this frequency be $\nu_p$, where the subscript is included since it will depend on the value of $p$. We can calculate this dependency explicitly: by definition, there will be $pN$ oscillators with frequency $-\nu_p \leq \omega \leq \nu_p$, leading to
\begin{equation}
\int_{-\nu_p}^{\nu_p} g(\omega)d\omega=p. 
\label{eq:gw_p}
\end{equation}
\noindent
With $g(\omega)=\frac{\gamma}{\pi(\omega^2+\gamma^2)}$, the integral in Eq.~(\ref{eq:gw_p}) reads 
\begin{equation}
2\int_{0}^{\nu_p} \frac{\gamma}{\pi}\frac{d\omega}{\omega^2+\gamma^2}
=\frac{2}{\pi}\tan^{-1}\frac{\nu_p}{\gamma},  
\end{equation}
which yields 
\begin{equation}
\nu_p = \gamma\tan \frac{p\pi}{2}.
\label{nu_p}
\end{equation}

\noindent
Now, we earlier remarked that the condition for an oscillator to be locked is $|\omega| \leq S $. Since on branch 1 only oscillators with $\xi_i > 0$ are locked, the condition to be on this branch becomes $0 \leq S \leq \nu_p$. The self-consistency 
equation \eqref{eq:self_consistency_eqn1} then becomes
\begin{eqnarray}
\label{eq:S}
S  &=& \langle\xi\cos\phi\rangle_{\rm{lock}}+\langle\xi\cos\phi\rangle_{\rm{drift}} \nonumber\\
   &=& \langle (+1) \cos\phi\rangle_{|\omega|\leq S}\\
   &=& R, \nonumber 
\end{eqnarray}
\noindent
where we used $\langle\cos\phi\rangle_{\rm{drift}}=0$ due to the symmetry about $\phi=\frac{\pi}{2}$.  So, for $S\leq \nu_p$, we get 
\begin{equation}
S=R=\langle\cos\phi\rangle_{\rm{lock}} = \int_{-S}^{S} d\omega~g(\omega)\sqrt{1-(\omega/S)^2}.
\label{eq:S_locked}
\end{equation}
Substituting $g(\omega)=\frac{\gamma}{\pi(\omega^2+\gamma^2)}$ in Eq.~(\ref{eq:S_locked}), we find
\begin{equation}
\int_{-1}^{1} dx \frac{\gamma}{\pi}\frac{S\sqrt{1-x^2}}{(Sx)^2+\gamma^2}=
\frac{\sqrt{S^2+\gamma^2}-\gamma}{S},~~{\rm{for}}~S \neq 0, 
\end{equation}
which gives
\begin{equation}
S=R=\sqrt{1-2\gamma}
\label{eq:S_gammac}
\end{equation}
for $\gamma\leq 1/2$. We note that Eq.~(\ref{eq:S_gammac}) is 
valid only for those $\gamma$ which 
satisfy $\sqrt{1-2\gamma} \leq \nu_p$. Using Eq.~\eqref{nu_p} for $\nu_p$, the 
critical $\gamma^*$ is given by

\begin{equation}
\gamma^*=\frac{\sec{\frac{p\pi}{2}}-1}{\tan^2\frac{p\pi}{2}},~~{\rm{for}}~0<p \leq 1.
\label{eq:gammastar}
\end{equation}

\noindent
This is value of the $\gamma$ that separates the two branches of $S$. Or more physically, the value of $\gamma$ below which oscillators with $\xi_i < 0$ start being locked along with those with $\xi_i > 0$.

In summary, the first branch of $S$ is given by Eq.~\eqref{eq:S_gammac}, which holds for $\gamma^* \leq \gamma \leq 1/2$. We draw two conclusions from this expression. The first is that there is a critical width $\gamma_c=\frac{1}{2}$, for all $p$, beyond which phase coherence disappears ($S=R=0$). Interestingly, this critical value does not depend on the value of $p$. The second is the scaling behavior at this critical point: $S\sim (\gamma_c-\gamma)^{\beta}$ with $\beta=\frac{1}{2}$, which is same as that of conventional mean-field systems including the traditional Kuramoto model (which is recovered by setting $p=1$).


\subsection{Branch 2}

The second branch of stationary states for $S$ is defined by $S>\nu_p$. This means both positively and negatively coupled oscillators are locked. But the drifters still do not contribute to the phase coherence. Hence $S$ is given by
\begin{eqnarray}
S &=&\langle \xi\cos\phi\rangle_{\rm{lock}}\nonumber\\
&=& \langle (+1) \cos\phi\rangle_{|\omega|\leq \nu_p} + \langle (-1) \cos\phi\rangle_{S\geq|\omega|>\nu_p} \nonumber\\
&=& 2\int_{0}^{\sin^{-1}(\nu_p/S)}\cos\phi~g(S\sin\phi) S\cos\phi~d\phi \nonumber\\
&-& 2\int_{\sin^{-1}(\nu_p/S)}^{\pi/2} \cos\phi~g(S\sin\phi) S\cos\phi~d\phi,
\end{eqnarray}
\noindent
where we used $\phi^*=\sin^{-1}(\nu_p/S)$ when $\omega=\nu_p$, and used 
$g(\omega)d\omega=g(S\sin\phi)S\cos\phi d\phi$ for the locked oscillators. The $S=0$ solution can be ignored, 
since we are assuming $S>\nu_p$.  
Thus, the second branch of partially locked states satisfies 
\begin{eqnarray}
\frac{1}{2} &=& 
\int_{0}^{\sin^{-1}(\nu_p/S)}\cos^2\phi~g(S\sin\phi)~d\phi \nonumber\\
&-& \int_{\sin^{-1}(\nu_p/S)}^{\pi/2}\cos^2\phi~g(S\sin\phi)~d\phi,
\label{eq:sc_S}
\end{eqnarray}
\noindent
where $\nu_p=\gamma \tan\frac{p\pi}{2}$ again. Evaluating the integrals yields the following equation:

\begin{eqnarray}
&&0=\pi(S^2-\gamma)+4\gamma\sin^{-1} \left(\frac{\gamma\tan \frac{p\pi}{2}}{S} \right)+ 2\sqrt{S^2+\gamma^2} \nonumber\\
&&\times \Bigg[\tan^{-1}\left(\frac{1}{\tan\frac{p\pi}{2}\sqrt{\frac{S^2+\gamma^2}{S^2-\gamma^2 \tan^2\frac{p\pi}{2}}}}\right)
\nonumber\\
&&~~~~~-\tan^{-1}\left(\tan\frac{p\pi}{2}\sqrt{\frac{S^2+\gamma^2}{S^2-\gamma^2 \tan^2 \frac{p\pi}{2}}}\right) \Bigg], \nonumber\\
\label{eq:sc2}
\end{eqnarray}

\noindent
which defines $S$ implicitly in terms of the parameters $p$ and $\gamma$. We were unable to solve Eq.~\eqref{eq:sc2} analytically, so instead we solved it numerically using Newton's method. We discuss the result in the next section.


\section{IV. Numerical Results}

To test our predictions for the branches of $S$ defined by Eq.~\eqref{eq:S_gammac} and Eq.~\eqref{eq:sc2}, we numerically integrated Eq.~\eqref{eq:model}. 
We used a fourth-order Runge-Kutta (RK4) method for $N = 12800$ oscillators with $\{ \phi_j(0) \}$ drawn uniformly at random. Our step size was $\delta t = 0.01$ for a total of $2 \times 10^6$ time steps. To avoid any transient behavior, we average the data over the final $1\times 10^6$ time steps.

\begin{figure}[!htpb]
       \includegraphics[width=0.95\linewidth]{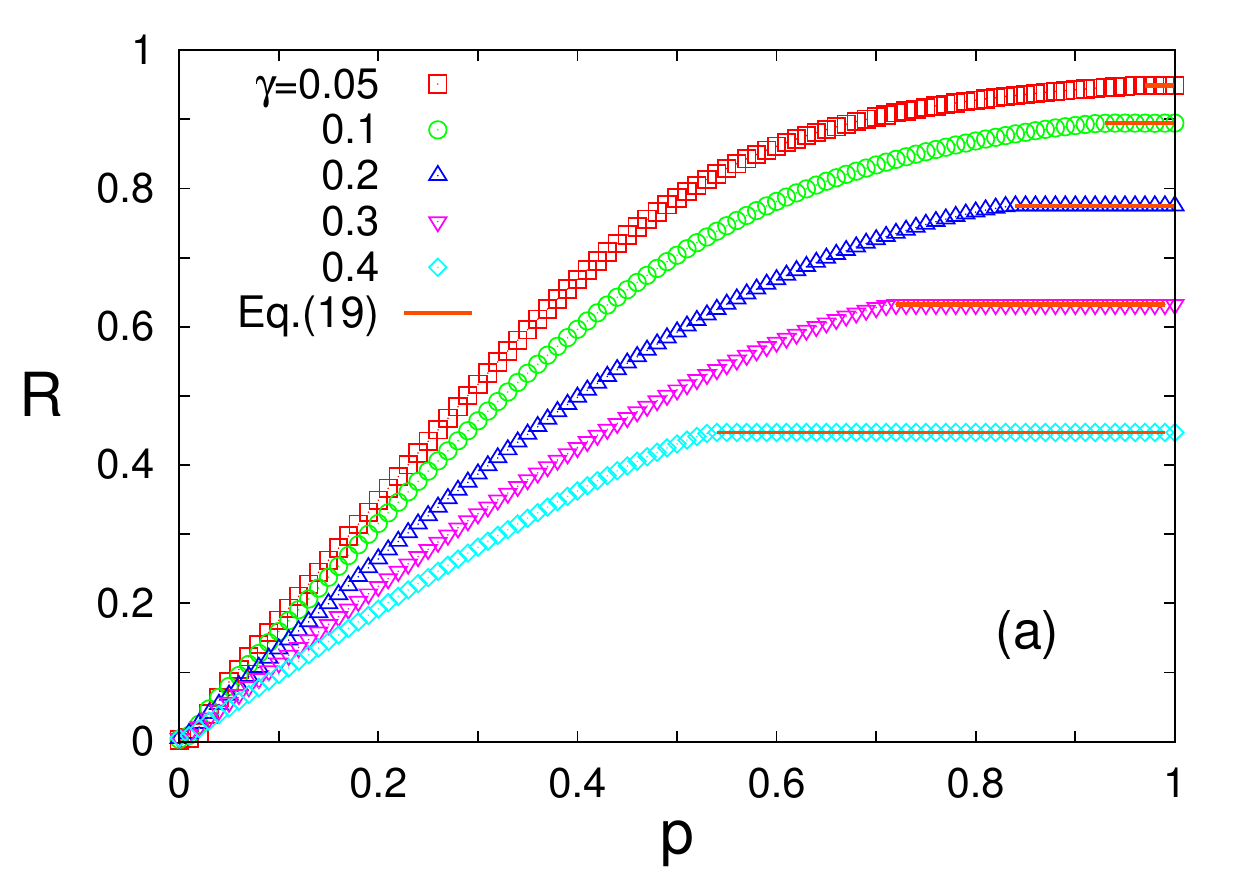}
       \includegraphics[width=0.95\linewidth]{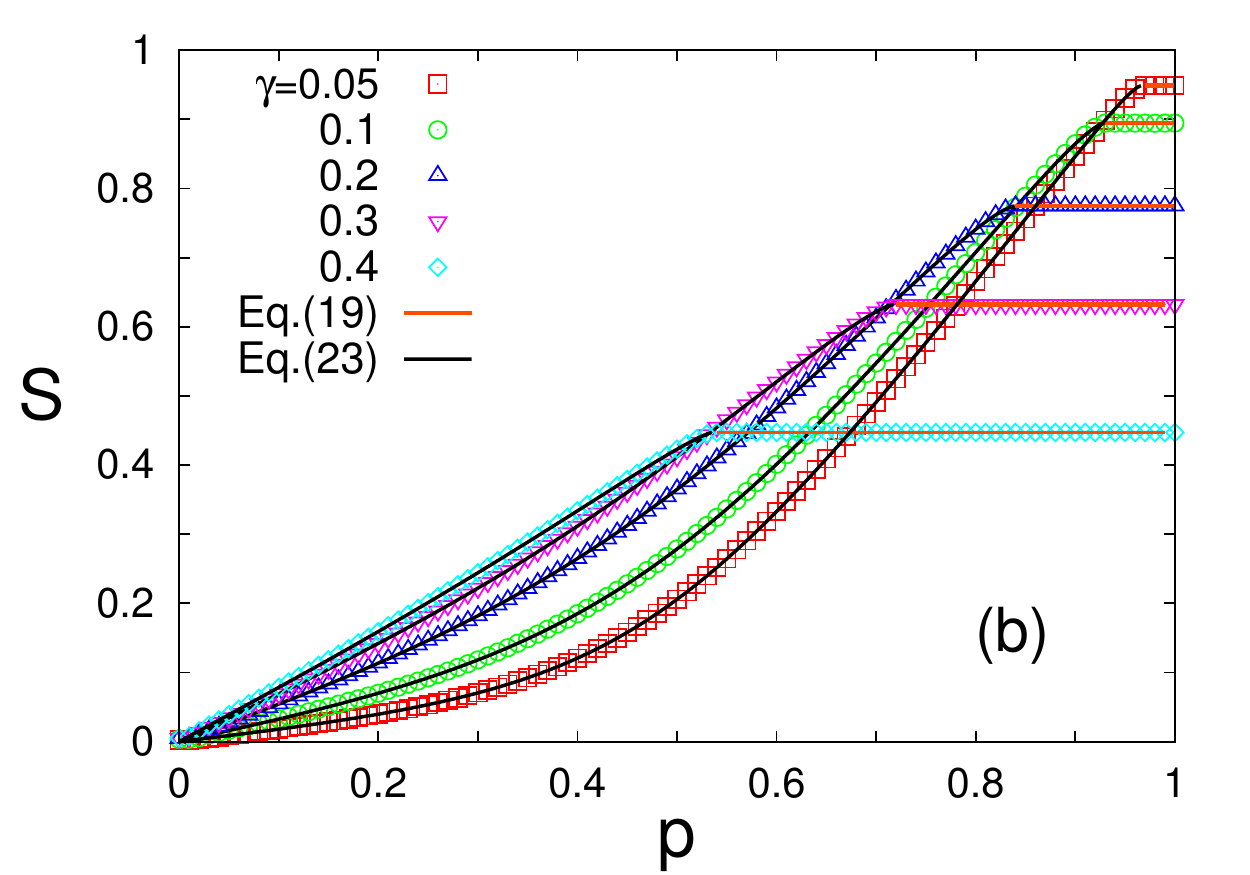}
        \caption{\label{fig:S_p_sym} 
(Color online) (a) Order parameter $R$ is plotted as a function of $p$ for 
various values of $\gamma$.  
Open symbols represent the data from the numerical simulations on Eq.~(\ref{eq:model}), 
with the system size $N=12800$.  The data have been averaged over 10 samples 
with different initial conditions $\{\phi_i(0)\}$, 
where the errors (not shown) are smaller than the symbol size.  
The red line shows the theoretical result from 
Eq.~{\eqref{eq:S_gammac}}, which predicts $R$ on branch 1 only.
(b) Order parameter $S$ is shown as a function of $p$ for 
various values of $\gamma$.  
Theoretical predictions from Eq.~(\ref{eq:S_gammac}) and (\ref{eq:sc2}) 
are also shown by the red solid line and black one, respectively.
The symbols represent the numerical simulation data, similar to (a).
}
\end{figure}

Figure~\ref{fig:S_p_sym} shows the behavior of $R$ and $S$ as a function of $p$ for various values of $\gamma$. The two branches of $R$ and $S$ are evident, and meet at a critical $p^*$. We earlier calculated this point in terms of a critical width $\gamma^*$ 
in Eq.~\eqref{eq:gammastar}. 
By rearrangement we can find $p^*$, defined implicitly via
\begin{equation}
1+\gamma \tan^2\frac{p^*\pi}{2} =\sec{\frac{p^*\pi}{2}}.
\label{eq:pstar}
\end{equation}

\noindent
When $p > p^*$, we are on branch 1. As predicted by Eq.~\eqref{eq:S_gammac}, both $S$ and $R$ are independent of $p$. Why is this? The reason is that on branch 1, only oscillators with $\xi_i > 0$ are locked, while those with both $\xi_i > 0$ and $\xi_i < 0$ are drifting. As we lower $p$, the number of oscillators with $\xi_i > 0$ in the drifting population get reduced. But the number in the locked population stays fixed, which means $S$ and $R$ remain fixed.

We arrive on the second branch when $p < p^*$. Here the situation is reversed: the drifting population consists purely of oscillators with $\xi_i < 0$, while the locked population contains those with both $\xi_i < 0$ and $\xi_i > 0$. The presence of locked oscillators with $\xi_i < 0$ lowers the magnitude of $S$, because they contribute negatively 
to the sum in Eq.~\eqref{eq:W}.  Consequently, the \textit{size} of the locked population is reduced, which in turn lowers both $S$ and $R$, which can been seen in Figure~\ref{fig:S_p_sym} for $p < p^*$. This is because the maximum frequency of the locked oscillators is given by $S = \omega_{max, locked}$; the lower $S$, the smaller the locked population.

\begin{figure}[!htpb]
       \includegraphics[width=0.9\linewidth]{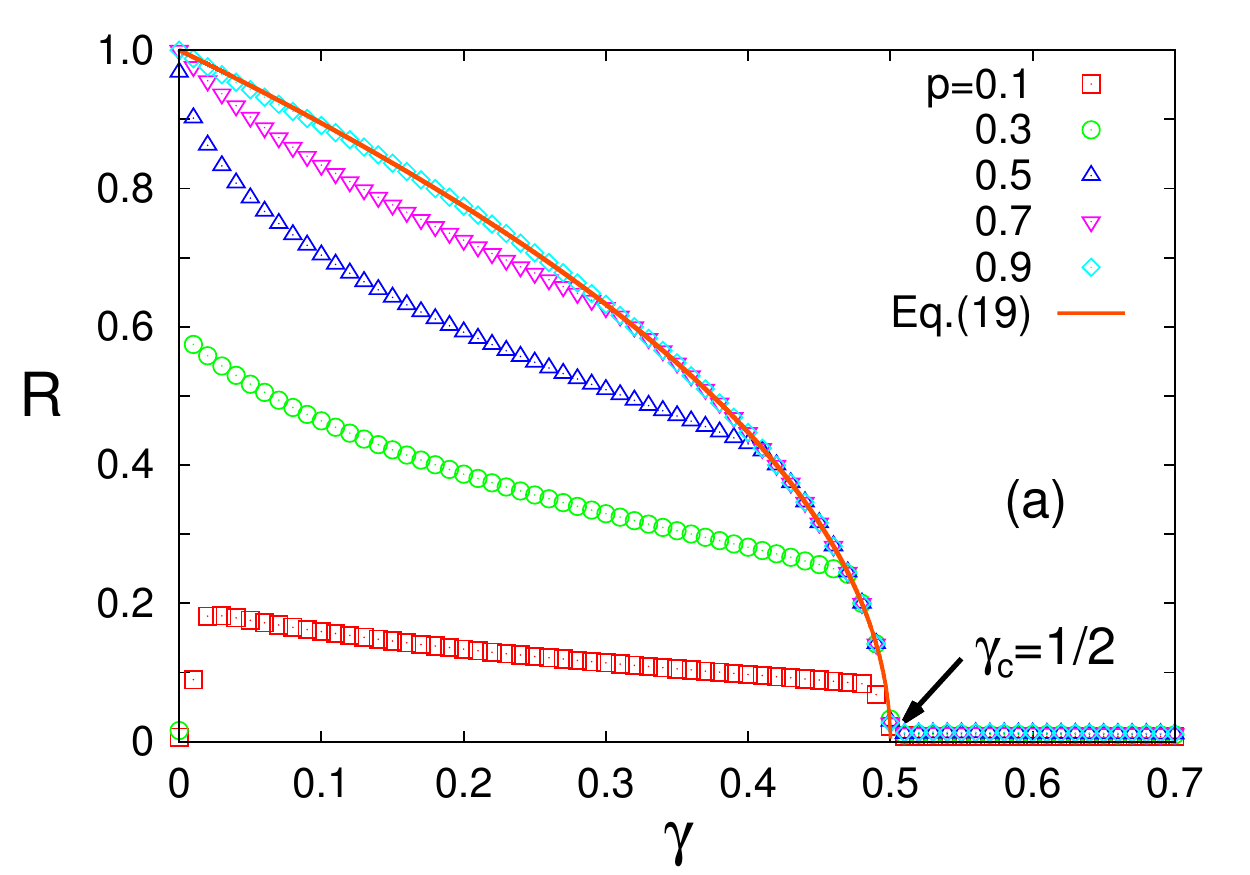}
       \includegraphics[width=0.9\linewidth]{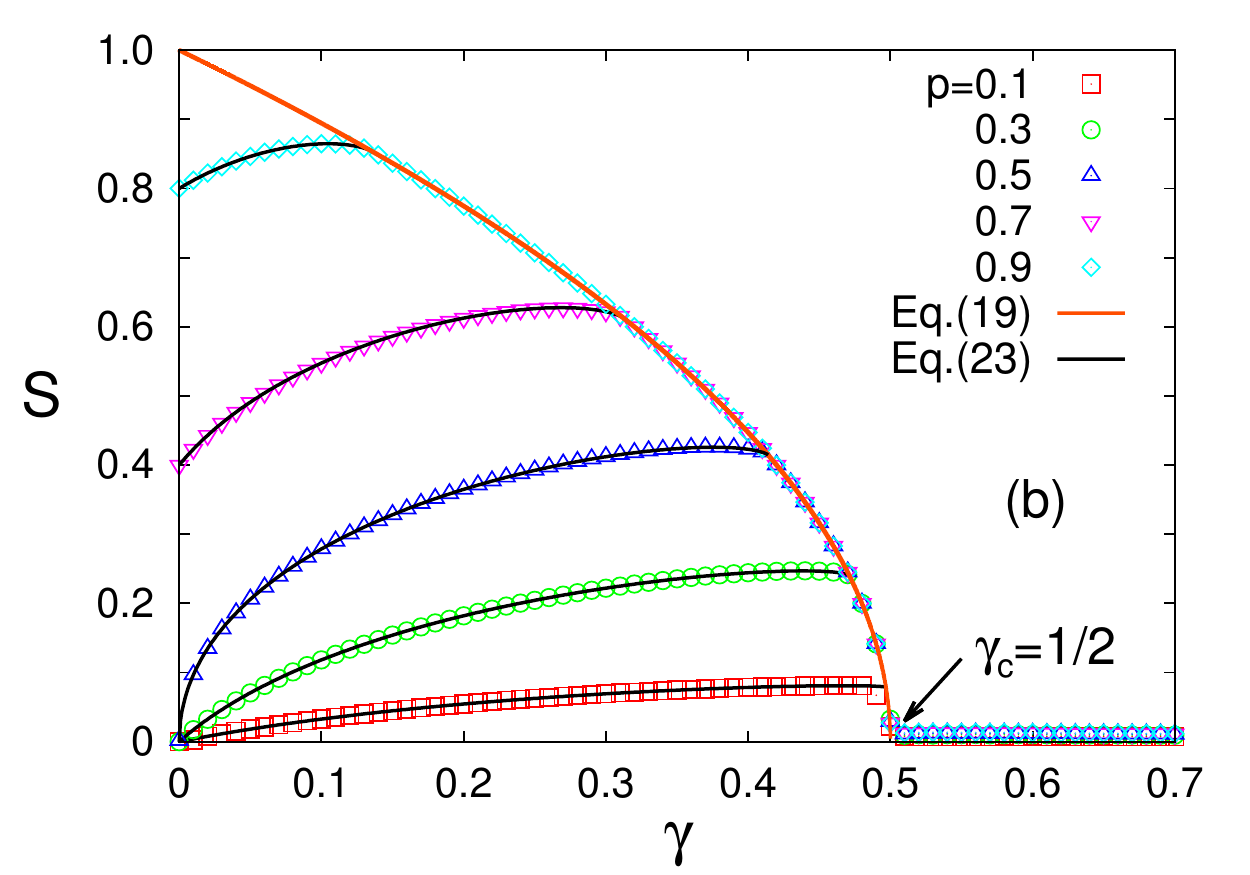}
        \caption{\label{fig:S_gamma_sym} 
(Color online) (a) Order parameter $R$ is plotted as a function of $\gamma$ for 
various values of $p$.  
Open symbols represent the data from the numerical simulations on Eq.~(\ref{eq:model}), 
with the system size $N=12800$.  The data have been averaged over 10 samples 
with different initial conditions $\{\phi_i(0)\}$, 
where the errors (not shown) are smaller than the symbol size.  
The red solid line shows the theoretical prediction from 
Eq.~(\ref{eq:S_gammac}) for $R$ on branch 1 only.
(b) Order parameter $S$ is shown as a function of $\gamma$ for 
various values of $p$. 
Theoretical predictions in Eq.~(\ref{eq:S_gammac}) and (\ref{eq:sc2}) 
are shown together by the red solid line and black one, respectively.
The symbols represent the numerical simulation data, similar to (a).
}
\end{figure}
In Figure~\ref{fig:S_gamma_sym}, we show $R$ and $S$  as a function of $\gamma$ for various values of $p$. As predicted by Eq.~\eqref{eq:S_gammac}, the values of $R$ and $S$ match the results from the 
traditional Kuramoto model on branch 1, 
which is defined for $\gamma^* \leq \gamma \leq \gamma_c$ (where $\gamma^*$ is given by Eq.~\eqref{eq:gammastar}, and $\gamma_c = 1/2$, which is derived from Eq.~\eqref{eq:S_gammac}). However on the branch 2, the results diverge, where now $S$ satisfies the implicit equation~\eqref{eq:sc2}.

\begin{figure}[!htpb]
       \includegraphics[width=0.9\linewidth]{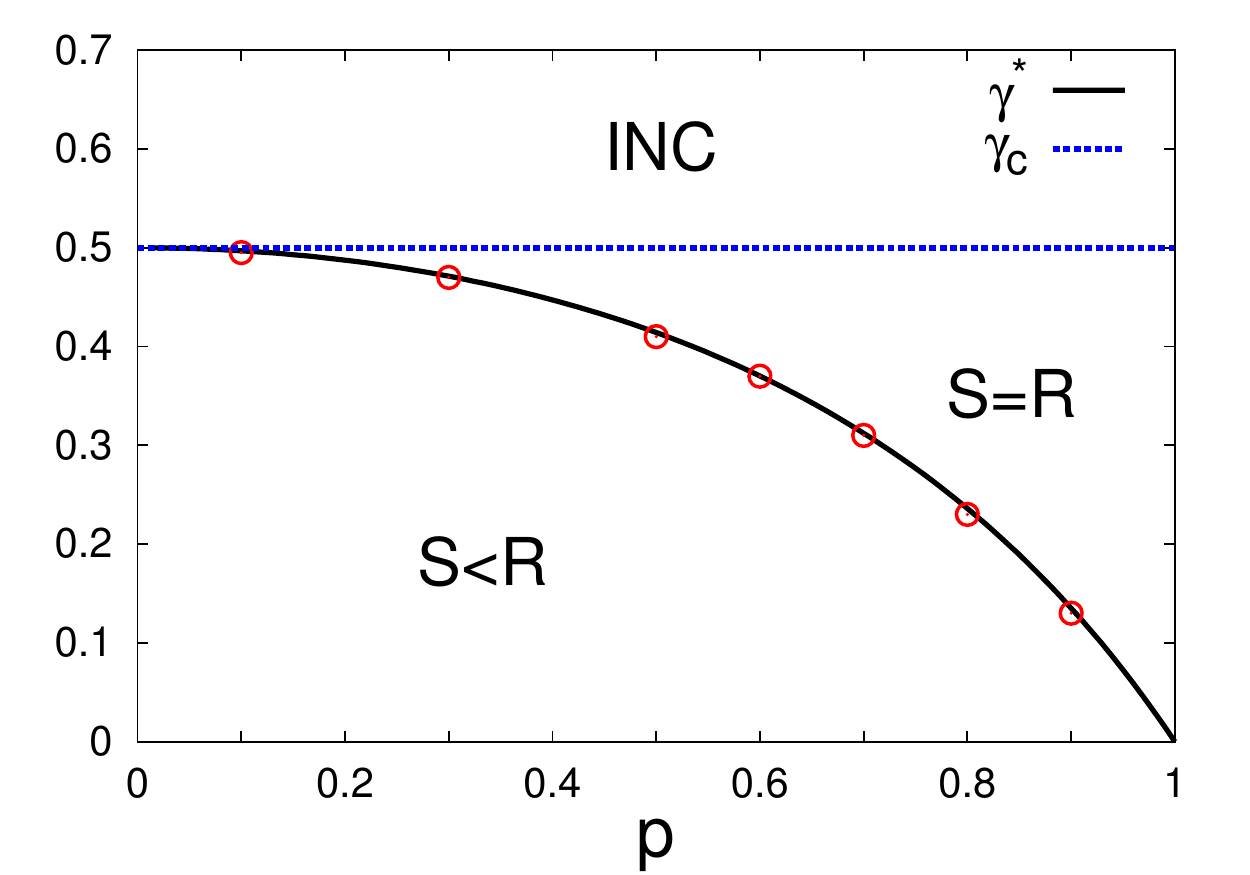}
       \caption{\label{fig:gammastar_sym} 
(Color online) 
The black solid line shows the teoretical prediction of $\gamma^*$ 
given by Eq.~(\ref{eq:gammastar}) as a function of $p$,
and the red open circles represent the numerical data obtained from the 
behavior of $R$ and $S$ shown in 
Figs.~{\ref{fig:S_p_sym} and \ref{fig:S_gamma_sym}}.
The blue dashed line displays $\gamma_c(=1/2)$ beyond which only the incoherent (INC) 
state with $S=R=0$ exists.
}
\end{figure}

Finally, in Fig.~\ref{fig:gammastar_sym} 
we show $\gamma^*$ versus $p$, where $\gamma^*$ is given by Eq.~(\ref{eq:gammastar}). Good agreement between theory and simulation is evident.

\section{V. Finite-size scaling of the order parameter}

In this section, we investigate the critical behavior near the transition point $\gamma=\gamma_c$ for large but finite values of $N$.  According to finite-size scaling theory~\cite{HHTP15} we expect the order parameter $S$ for a given value of $p$ to satisfy
\begin{equation}
S(\gamma, N; p)=N^{-\beta/\bar\nu}f(\epsilon N^{1/\bar\nu})
\label{eq:S_FSS}
\end{equation}
in the critical region, where $\epsilon=\gamma-\gamma_c$ and $f(x)$ is a scaling function having the asymptotic properties
\begin{equation}
f(x) =
\left\{
\begin{array}{ll}
\rm{const}, & x=0,\\
(-x)^{\beta}, & x \ll -1,\\
x^{\beta-\bar\nu/2}, & x \gg 1. \\
\end{array}
\right.
\label{eq:f}
\end{equation}
\noindent 
The exponent $\bar\nu$ is the finite-size scaling exponent, and $\beta$ is the order-parameter exponent, i.e., $\beta=1/2$ from the critical behavior of the order parameter: $S\sim (\gamma_c-\gamma)^{1/2}$.  
The size dependence of the order parameter $S$ at the transition ($\epsilon=0$) allows us to estimate the decay exponent $\beta/\bar\nu$. Note that we expect the conventional order parameter $R$ to qualitatively have the same scaling behavior as $S$.

\begin{figure}[!htpb]
       \includegraphics[width=0.95\linewidth]{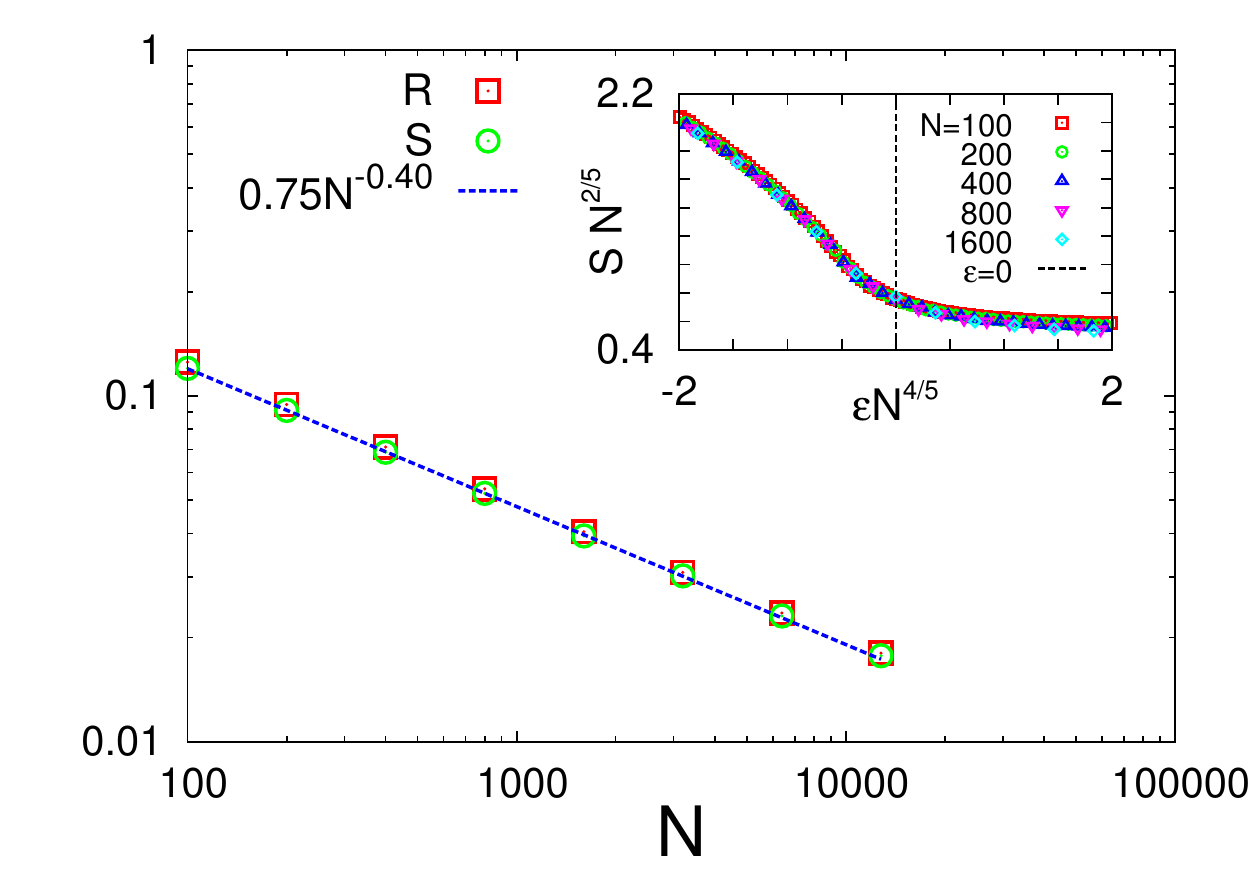}
        \caption{\label{fig:S_N_g05_sym} 
(Color online) Critical decay of the order parameters $R$ and $S$ at 
transition $\gamma=\gamma_c(=1/2)$ is plotted as a function of the system size 
$N$ in a log-log plot.  The red open boxes and the green open circles 
represent $R$ and $S$, respectively, where the numerical data of $R$ and 
$S$ have almost the same values. The data are averaged over 10 samples, where 
the errors are symbol size (not shown).  The slope of the straight line is 
given by -0.4, which means $\beta/\bar\nu=2/5$.  Inset: Data collapsing of 
$S$ for various size $N$ is shown, where $\beta/\bar\nu=2/5$ and $\bar\nu=5/4$ 
have been used at $\gamma=1/2$ for $p=1/2$.}
\end{figure}

To test these predictions, we numerically measured $S$ at $\epsilon=0$ (i.e., at the critical width  $\gamma=\gamma_c$), for various system sizes.  Figure~\ref{fig:S_N_g05_sym} shows the critical decay of the order parameters $R$ and $S$ at the transition point, $\gamma=1/2$ for $p=1/2$, where its slope gives $\beta/\bar\nu=2/5$ for both $R$ and $S$.  
Since $\beta=1/2$ from the self-consistency analysis, we deduce $\bar\nu=5/4$.  The scaling plot of the order parameter $S$ for various system sizes $N$ is shown in the inset of Fig.~\ref{fig:S_N_g05_sym}, which displays a good collapse of the data. The parameter values we used were $\beta/\bar\nu=2/5$ and $\bar\nu=5/4$ at $p=1/2$ and $\gamma_c=1/2$.

We have also investigated the finite-size scaling of the order parameters at the other values of $p$ such as $p=3/10$ and $p=7/10$, and obtained the same result: $\bar\nu=5/4$. This value for $\bar\nu$ is the same as that obtained for the 
traditional Kuramoto model 
with deterministically chosen natural frequencies ~\cite{HHTP15}. 
That is, choosing $\{\omega_i \}$ and $\{\xi_i \}$ 
according to the deterministic procedure \eqref{eq:regLorw} and 
\eqref{eq:xi_sym}, respectively, gives the same finite scaling exponent.
Lastly, it is also interesting to note that the system shows the same finite-size scaling exponent $\bar\nu$ as that for the Kuramoto model even for the case that $p<1$. 

\section{VI. Dynamic fluctuation of the order parameter}
In this section, we investigate the dynamic fluctuation of the order parameter.  
Although we assumed that the order parameters $S$ and $R$ were time-independent in the infinite-$N$ limit, those quantities will exhibit fluctuations for finite $N$. The dynamic fluctuation $\chi_A$ of an order parameter $A(t)$, which for us could be either $R(t)$ or $S(t)$, is defined~\cite{HHTP15} as
\begin{equation}
\chi_A (\bar\gamma, N) \equiv N\langle\langle A^2\rangle_t - {\langle A \rangle_t}^2 \rangle,
\label{eq:chiS_chiR}
\end{equation}
\noindent
where $\langle\cdots\rangle_t$ and $\langle\cdots \rangle$ represent the time 
average and the  sample average, respectively.  Here, one sample means one 
configuration with an initial condition $\{\phi_i(0)\}$.  We expect that the 
critical behavior of $\chi_A$ will be given by
\begin{equation}
\chi_A =
\left\{
\begin{array}{ll}
(-\epsilon)^{-\bar\gamma}, & \epsilon < 0,\\
\epsilon^{-\bar\gamma^{\prime}}, & \epsilon > 0 \\
\end{array}
\right.
\label{eq:def_chiS}
\end{equation}
\noindent
in the thermodynamic limit $N\rightarrow \infty$.

The two exponents $\bar\gamma$ and $\bar\gamma^{\prime}$ characterize the diverging behavior of $\chi$ in the supercritical $(\epsilon<0)$ and subcritical $(\epsilon > 0)$ regions, respectively. For most homogeneous systems, scaling is controlled by a single exponent: $\bar\gamma=\bar\gamma^{\prime}$~\cite{HHTP15}. The corresponding finite-size scaling is then given by
\begin{equation}
\chi_A (\epsilon, N) = N^{\bar\gamma/\bar\nu} h(\epsilon N^{1/\bar\nu}),
\label{eq:chiS}
\end{equation}
\noindent
where the scaling function $h(x)$ again has the form
\begin{equation}
h(x) =
\left\{
\begin{array}{ll}
\rm{const}, & x=0,\\
(-x)^{-\bar\gamma}, & x \ll -1, \\
x^{-\bar\gamma}, & x \gg 1. \\
\end{array}
\right.
\label{eq:h}
\end{equation}

\begin{figure}[!htpb]
       \includegraphics[width=0.95\linewidth]{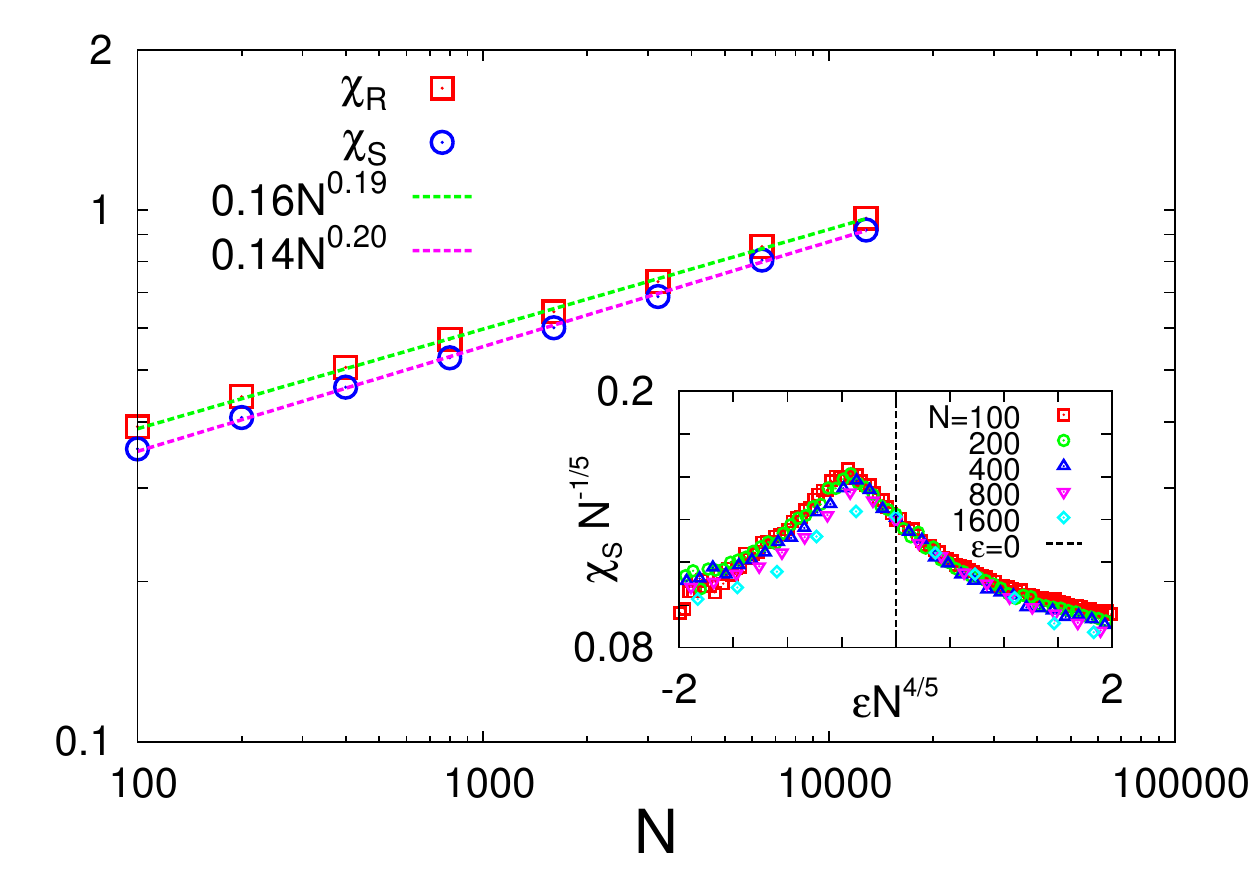}
        \caption{  \label{fig:chiS_N_g05_sym} 
(Color online) Critical increase of $\chi_R$ and $\chi_S$ at 
$\gamma=\gamma_c(=1/2)$ is shown as a function of $N$ in a log-log plot, 
where $p=1/2$ is chosen for convenience.  
The slopes of the two straight lines are 0.19 and 0.20, respectively, 
displaying a good agreement with $\gamma/\bar\nu \approx 1/5$.  Inset: Scaling plot of 
$\chi_S$ for various sizes $N$ is shown, where $\beta/\bar\nu=2/5$ and 
$\bar\nu=5/4$ have been used at $\gamma=1/2$ for $p=1/2$.}
\end{figure}

Figure~\ref{fig:chiS_N_g05_sym} shows the critical increase of $\chi_R$ and $\chi_S$ at the transition $(\gamma_c=1/2)$ for $p=1/2$, where the slopes of the straight lines are given by 0.19 and 0.20, respectively. 
This implies $\bar\gamma/\bar\nu \approx 1/5$. Substituting $\bar\nu=5/4$, we find that $\bar\gamma = 1/4$. The inset of Fig.~\ref{fig:chiS_N_g05_sym} shows the scaling plot of $\chi_S$, with $\bar\gamma/\bar\nu=1/5$ and $\bar\nu=5/4$. A good collapse of the data is evident.

We also investigated the dynamic fluctuation of $S$ at the other values of $p$ such as $p=3/10$ and $7/10$, where we found the same result: $\bar\gamma \approx {\bar\gamma}^{\prime} \approx 1/4$ for $p=7/10$. However, for $p < 3/10$, the numerical results are too inaccurate to confidently determine the value of $\bar\nu$. More substantial numerical experiments are required to resolve this issue, which we leave for future work.

The critical exponents we found for $p=1/2$ and $p=7/10$ are 
\begin{equation}
\beta=1/2, ~~\bar\nu=5/4, ~~\bar\gamma=\bar\gamma^{\prime}\approx 1/4,
\label{eq:criticalexponents}
\end{equation}
\noindent
which shows that the hyperscaling relation
\begin{equation}
\bar\gamma=\bar\nu-2\beta
\end{equation}
\noindent
holds for our model \eqref{eq:model} with symmetrically correlated disorder. These results tell us that our model belongs to the same universality class as the simpler Kuramoto model with deterministic disorder in the frequencies, but with constant positive coupling~\cite{HHTP15}.

\section{VII. Summary}

We have studied a mean-field model of coupled phase oscillators with quenched, correlated disorder -- specifically,  when the natural frequencies are ``symmetrically'' correlated with the couplings. We found that these correlations enhanced the synchronizability of the system, in the sense that the partially locked state could occur for any $p > 0$ (where $p$ is the fraction of positively coupled oscillators). We further found that this state only occurs when the width of the frequency distribution is lower than a critical value $(\gamma < \gamma_c)$, irrespective of $p$. Interestingly, this threshold is found to be same as that of the 
traditional Kuramoto model (which is recovered from our model when $p=1$). 

We also explored the finite-size scaling behavior as well as the dynamic fluctuation of the order parameter. We found the system belongs to the same universality class as the Kuramoto model with 
deterministically chosen natural frequencies, even for $p<1$.   
Curiously, when randomness comes into the system, the finite-size scaling and dynamic fluctuations seem to show different behavior from the case with 
deterministically chosen correlated disorder, which requires further study~\cite{ref:future}. 

There are many variants of the Kuramoto model in which we could 
further study the effects of correlated disorder. 
One example is a model closely related to ours, studied in \cite{HS11}. The difference between the two models is that the coupling $\xi_i$ is \textit{outside} the sum in \cite{HS11}, whereas it appears \textit{inside} the sum in Eq.~\eqref{eq:model}. 
This key difference leads to qualitatively new states, such as the traveling wave 
state and the $\pi$ state. 
It would be interesting to see how correlations influence these more exotic states.

\section{acknowledgments}
This research was supported by NRF Grant No. 2015R1D1A3A01016345 (to H.H.) and 
NSF grant DMS-1513179 and CCF-1522054 (to S.H.S).


\end{document}